\documentclass[aps,pra,reprint,superscriptaddress,amsmath,amssymb,longbibliography]{revtex4-2}

\usepackage{graphicx}% Include figure files
\usepackage{dcolumn}% Align table columns on decimal point
\usepackage{bm}% bold math
\usepackage{todonotes}
\usepackage{hyperref}% add hypertext capabilities

\begin{document}

\title{Signatures of the BCS-BEC crossover in the yrast spectra of  Fermi quantum rings}

\author{Ulrich Ebling}
\email{uebling97@gmail.com}
\affiliation{Centre for Theoretical Chemistry and Physics,
New Zealand Institute for Advanced Study,  Massey University, Private Bag 902104, North Shore, Auckland 0745, New Zealand}
\affiliation{Dodd-Walls Centre for Photonic and Quantum Technologies, PO Box 56, Dunedin 9056, New Zealand}
\author{Ali Alavi}
\affiliation{
 Max Planck Institute for Solid State Research, Heisenbergstra{\ss}e 1,
70569 Stuttgart, Germany
}%
 \affiliation{Department of Chemistry, University of Cambridge,
Lensfield Road, Cambridge, CB2 1EW, United Kingdom}
\author{Joachim Brand}
\affiliation{Centre for Theoretical Chemistry and Physics,
New Zealand Institute for Advanced Study,  Massey University, Private Bag 902104, North Shore, Auckland 0745, New Zealand}
\affiliation{Dodd-Walls Centre for Photonic and Quantum Technologies, PO Box 56, Dunedin 9056, New Zealand}
\affiliation{
 Max Planck Institute for Solid State Research, Heisenbergstra{\ss}e 1,
70569 Stuttgart, Germany
}%

\date{\today}

\begin{abstract}
We study properties of the lowest energy states at non-zero total momentum (yrast states) of the Hubbard model for spin-$\frac{1}{2}$ fermions in the quantum ring configuration with attractive on-site interaction at low density. In the one-dimensional (1D) case we solve the Hubbard model using the Bethe ansatz, while for the crossover into the 2D regime we use the Full-Configuration-Interaction Quantum Monte-Carlo method (FCIQMC) to obtain the yrast states for the spin-balanced Fermi system. We show how the yrast excitation spectrum changes from the 1D to the 2D regime and how pairing affects the yrast spectra. We also find signatures of fragmented condensation for certain yrast states usually associated with dark solitons.
\end{abstract}

%\keywords{Suggested keywords}%Use showkeys class option if keyword
                              %display desired
\maketitle

%\tableofcontents

\section{Introduction}
The crossover from a fermionic superfluid of weakly-bound Cooper pairs (BCS regime) to a Bose-Einstein condensate (BEC) of strongly-bound dimers is a paradigmatic quantum many-body problem \cite{Eagles1969,Leggett1980a,Zwerger2012,Levinsen2015}.
% In the crossover regime our
Our understanding of this problem is still limited, as strong quantum correlations and the absence of a small parameter pose severe challenges for theoretical and computational approaches.
% the problem still poses a severe challenge for theoretical and computational approaches due to strong quantum correlations and the absence of a small parameter that would allow the application of perturbation theory.
While bulk systems have been studied extensively in recent years using theory  \cite{Zwerger2012,Giorgini2008,Parish2014} and experiments with quantum gases \cite{Regal2004,Zwierlein2004,Chin2004,Bourdel2004,Kinast2004,Ketterle2008,Ku2012,Parish2014,Carcy2019,Mukherjee2019},
%\cite{Regal2004,Chin2004,Ketterle2008,},
the advent of quantum gas microscopes \cite{Bakr,Cheuk2015,Haller2015,Brown2020} and micro traps \cite{Grunzweig2010b,Serwane2011,Wenz2013a,Reynolds2020} has opened up the opportunity to  experiment with
%perform detailed experimental studies on
systems that are small enough to perform exact numerical calculations on.

Of particular interest are ring configurations, where translational invariance along one spatial dimension makes (angular) momentum a good quantum number. This allows for the study of yrast states, which are defined as the lowest energy state at given value of the total momentum. Yrast states in a bosonic superfluid are intimately connected \cite{Kulish1976,Kanamoto2008,Kanamoto2010,Jackson2011,Fialko2012,Sato2012a,Syrwid2015,Shamailov2019} to localized nonlinear waves known as dark solitons \cite{Tsuzuki1971}.
% In bosonic superfluids, specifically, i
It was shown that measuring the position of all bosons in an yrast state reveals a dark-soliton-like particle depletion \cite{Syrwid2015},
%While yrast states are fragmented quantum condensates, breaking translational invariance ,
and that wave-packet-like superpositions of yrast states emulate the behavior of classical dark solitons \cite{Shamailov2019}.
While the yrast states are fragmented quantum condensates, breaking the translational symmetry restores single condensation and classical soliton features \cite{Fialko2012}.
Dark solitons in Fermi superfluids  have been identified in experiments \cite{Yefsah2013,Ku2014,Ku2016}, but many predictions from mean-field theory have not yet been tested \cite{Antezza2007,Liao11pr:FermiSolitons,Scott2011,Spuntarelli2011,Scott2012,Cetoli2013,Efimkin2014}. Moreover, there is an intriguing connection \cite{Yoshida2007,Lutchyn2011}  between dark solitons and the predicted Fulde-Ferrell-Larkin-Ovchinnikov (FFLO) phase  of imbalanced superfluids \cite{Fulde1964,Larkin1964}. Dispersion relations of yrast states were analyzed in the context of dark soliton physics in the Yang-Gaudin model, a one-dimensional (1D) Bethe-ansatz solvable model of a fermionic superfluid in Ref.~\cite{Shamailov2016}. An overview of  computational studies of quantum rings can be found in the recent review literature \cite{Viefers2004,Manninen2012,Fomin2018}.

Beyond the purely one-dimensional models of quantum rings, a second spatial dimension can be added by considering stripe- or ladder-type lattice configurations as it is done in this work. In Refs.~\cite{Viefers2004,Manninen2012} these are referred to as quasi-1D geometries. While the Bethe ansatz is unavailable for such models and mean-field theory is not valid in the strongly-correlated regime, direct numerical simulation is very challenging due to the fact that Hilbert space size increases exponentially both with particle number and the number of lattice sites. Quantum Monte Carlo (QMC) simulations are still possible, although  the \emph{fermion sign problem} \cite{Troyer2005,Wu2005} provides a challenge for the simulation of fermionic many-body problems.

%
% in this situation and many different methods  are available.
%%The study of the quantum many-body problem remains one of the central challenges of physics. Quantum correlations and the exponentially increasing Hilbert space size required for an accurate description complicate numerical approaches in cases where exact solutions such as the Bethe ansatz are unavailable and mean-field theory is not valid. For strongly correlated systems, Quantum Monte Carlo (QMC) simulations are a widespread tool, with many different methods available.
%In particular, strongly interacting Fermi systems, where QMC suffers from the \emph{fermion sign problem} \cite{Troyer2005,Wu2005}, prove a challenge for these types of approaches. One example of this are Fermi superfluids in ultracold atomic gases, which have been achieved in experiments and offer a high degree of control over system parameters such as interactions and the system geometry.

QMC methods relevant in the field of ultracold quantum gases each have their own strengths and weaknesses. Diffusion Monte Carlo has no basis set dependence but either converges to a bosonic ground state or requires node-fixing, which introduces an approximation \cite{Foulkes2001}. Auxiliary-field QMC is sign-problem free for the attractive balanced Hubbard model \cite{Carlson2011}, but the Hubbard-Stratonovich transformation involved in this method breaks symmetries of the Hamiltonian and thus does not allow for the study of yrast states.
Recent work has suggested a solution  \cite{Motta2019} but it has yet to be seen whether the method can be implemented efficiently.
%Recent work has presented a solution to this which however appears to be highly non-trivial to implement \cite{Motta2019}.
Determinant Monte Carlo and related methods can handle finite temperature  and extrapolate to zero temperature, but they fix the chemical potential instead of particle number \cite{Varney2009,Wolak2010,Wolak2012,Mitra2018}. All these existing methods have in common that they can study overall ground-state properties while it is not possible to study yrast states, because the total momentum cannot be easily constrained.
% it is impossible or very complicated to restrict the simulation to a subset of Hilbert space which corresponds to a specific symmetry.

For this work we use Full-Configuration-Interaction Quantum Monte Carlo (FCIQMC), a method originally developed for strongly correlated electrons in the context of quantum chemistry \cite{Booth2009,Cleland2010}. FCIQMC has been applied with great success to a large number of problems in this field \cite{Booth2013,Cleland2012}, and recently to ultracold atoms \cite{Jeszenszki2020,Yang2020}. This method can find the ground state energy and many-body wave function in a Fermi system by expanding the wave function into a set of Slater determinants. A stochastic version of exact diagonalization of the Hamiltonian in this basis is achieved by simulating the dynamics of a walker population in Slater determinant space. FCIQMC mitigates the sign problem by walker anihilation
%addresses the sign problem
to a certain degree
but does not eliminate it \cite{Spencer2012}.
By performing a stochastic projection to the ground state of a Hamiltonian in a given Fock basis directly, i.e. without resorting to a Hubbard-Stratonovich transformation, it is easy to respect symmetries of the Hamiltonian. In particular, it is possible to obtain energies and observables from yrast states by ground state projection in a plane-wave basis because the FCIQMC algorithm conserves total momentum if the Hamiltonian does.
% In this paper, we utilize a feature of FCIQMC which makes it very easy compared to other QMC methods to find the lowest energy eigenstate corresponding to a symmetry that commuted with the Hamiltonian, in out case, the momentum operator. We calculate not only ground state solutions, but also the lowest energy eigenstates with finite total momentum, called yrast states.
With the FCI method taking into account all correlations in the system, we probe the BCS-BEC crossover from the non-interacting to the strongly attractive regime in the Hubbard model.
%
%In this work we consider a Hubbard model for spin-$\frac{1}{2}$ fermions with periodic boundary conditions in one dimension and filling factor much smaller than one. There is a body of literature on such systems, which are often called quantum rings, e.g.~see the review papers \cite{Viefers2004,Manninen2012}. \textbf{[Need to say more about previous literature on quantum rings and challenges going to 2D.]}
%
%Yrast states are related to quantum dark solitons, non-linear wave phenomena characterized by a localized density depletion stabilized by non-linear effects. Both repulsive bosons in 1D and attractive superfluid fermions have yrast energy dispersions which resemble that of dark soliton solutions of the Gross-Pitaevskii equation (cite). Previous work demonstrated that dark solitons can be constructed from superpositions of states located around the maxima of the yrast dispersion of the Lieb-Liniger model \cite{Jackson2011,Shamailov2016,Shamailov2019}. While solitons can be described using mean-field theory such as the Gross-Pitaevskii or Bogoliubov-de Gennes equations and are typical features of a single Bose-Einstein condensate, yrast states are characterized by fragmentation and the related dark soliton only appears by breaking the symmetry of the system \cite{Fialko2012}. In this paper, we show that in the Hubbard model, yrast states correspond to fragmented condensation which can be observed in the momentum-space pair density.

In this paper, we present a QMC study of yrast states in the Hubbard model for spin-$\frac{1}{2}$ fermions. Using a filling factor much smaller than one, this system resembles a continuum superfluid with the difference that momentum is replaced by lattice momentum. We study the crossover from 1D to 2D geometry in the case of attractive on-site interactions. For 1D Hubbard chains, we obtain exact results using the Bethe ansatz and compare them to QMC results. Then we use FCIQMC to investigate the crossover into the 2D geometry by increasing the number of sites in the transverse direction. We find that in 2D, the shape of the yrast dispersion changes considerably, but that for increasing interaction strength, the more typical shape of the 1D spectrum is restored, which indicates soliton-like physics and is a signature of the transition into the superfluid regime. We investigate in more detail the behavior of the local minima of the yrast spectrum at the so-called ``umklapp'' points where sufficient quasi-momentum is added to boost either all or half of the constituent fermions by one unit in order to form a ring current.
% ("umklapp points"), whose excitation energies  exhibit mesoscopic behavior before in 2D systems for weak interactions.
We find signatures of the transition from non-interacting Fermi gas to a paired superfluid and of the BCS to BEC crossover in the excitation energy and in the pair correlation functions for the first half and full umklapp points.
%By calculating the pair correlation functions we observe the BEC-BCS transition for the first Umklapp point.
Last, we focus on the yrast states around the maxima of the dispersion, which are related to dark solitons. We calculate the inertial mass of possible solitons for different system sizes and interaction strengths. We find an increase of the inertial mass by a factor of 2 when the transverse dimension is large enough for the system to be considered truly 2D, which is indicative of a transition from dark soliton to a solitonic vortex \cite{Brand2001,Brand2002}. From mean-field and basic hydrodynamic theory,  in the 1D to 2D crossover dark solitons are replaced as stable yrast excitations by solitonic vortices \cite{Brand2002,Mateo2014}, or vortex pairs \cite{Cetoli2013,VanAlphen2019}, which have larger inertial mass \cite{Ku2014,Mateo2015a,Toikka2016}.  By looking at the pair densities of these yrast states, we find that fragmented condensation into more than one momentum state takes place, as expected for superfluid yrast states  \cite{Fialko2012}.

This paper is organized as follows:
After introducing the model in Sec.~\ref{sec:system} and the FCIQMC approach in Sec.~\ref{sec:fciqmc}, we discuss yrast spectra of a 1D Hubbard chain obtained by the Bethe ansatz in Sec.~\ref{sec:bethe}. The energies and spin correlation functions for the umklapp points  in the 1D to 2D crossover are discussed in Sec.~\ref{sec:umklapp} before analysing the physics of the maxima of the yrast dispersion by computing their effective mass and momentum-space pair densities in Sec.~\ref{sec:maxima}, and drawing conclusions in Sec.~\ref{sec:conclusions}.

\section{System} \label{sec:system}
To study yrast states in an ultracold fermionic superfluid, we use the Hubbard model in the regime of low densities. In this regime, the Hubbard model approximates a discretized free space. The correspondence to free space becomes exact in the low-density limit.
%can be understood as a means of discretizing free space, or in other words as a basis set expansion where the cutoff is given by the number of lattice sites inside a box with periodic boundary conditions.
We focus on the crossover between the 1D and 2D geometry, therefore our Hubbard model corresponds to a rectangular lattice with $L\times W$ sites, where $1\le W<L$, but with the same lattice spacing $\alpha$ in both dimensions. We are using periodic boundary conditions in both directions, and thus our systems has the topology of a torus. The Hubbard Hamiltonian in momentum representation is
 \begin{equation}
 \label{eq:hubbard}
 H=\sum_{\vec k,s}\epsilon_{\vec k} c_{\vec k s}^\dagger c_{\vec k s}+\frac{U}{LW}\sum_{\vec k_1,\vec k_2,\vec k_3} c_{\vec k_1 \uparrow}^\dagger c_{\vec k_2 \downarrow}^\dagger c_{\vec k_3 \downarrow} c_{\vec k_1+\vec k_2-\vec k_3 \uparrow},
 \end{equation}
where the operators $c^{(\dagger)}_{\vec k,s}$ create (annihilate) a fermion with lattice momentum $\hbar \vec k$ and spin $s$. Wave vectors can take on values $\vec k=(k_x,k_y)=2\pi(n_x/\alpha L,n_y/\alpha W)$, where $n_x,n_y$ are integers. The single-particle dispersion for the Hubbard model is given by
\begin{equation}
\label{eq:Hubbard_dispersion}
    \epsilon_{\vec k}=2t(2-\cos(k_x\alpha)-\cos(k_y\alpha))
\end{equation}
where $t$ is the hopping amplitude and $U<0$ is the interaction parameter. In the low density regime, mostly the low-lying momentum states are occupied where the dispersion relation (\ref{eq:Hubbard_dispersion}) is approximately parabolic. Thus the Hubbard model approximates a continuum Fermi gas.

Throughout this paper we present results obtained for a particle number of $N=10$, with 5 fermions in each spin state, and a lattice length of $L=21$, while the width $W$ varies from 1 to 11. Energies will be given in units of hopping amplitude $t$ and momenta in longitudinal lattice units $P_0=\frac{2\pi\hbar}{\alpha L}$.

\section{FCIQMC}\label{sec:fciqmc}
FCIQMC  is a numerical method originally created in the context of strongly correlated electron systems and quantum chemistry. It can find ground states of fermionic many-body systems by expanding the many-body wave function in terms of Slater determinants
\begin{equation}
    |\Psi\rangle=\sum_i C_i |D_i\rangle,
\end{equation}
which in our case are of the form
\begin{equation}
    |D\rangle=c_{\vec k_1,s_1}^\dagger\ldots c_{\vec k_N, s_N}^\dagger|\text{vac}\rangle.
\end{equation}
FCIQMC then obtains the expansion coefficients $C_i$ by using a stochastic population dynamics approach to solve the imaginary-time Schr\"odinger equation. The automatic antisymmetrization of the wave function by expanding it in Slater determinants ensures that unlike Diffusion Monte-Carlo, FCIQMC always finds a fermionic wave function. FCIQMC's capacity for overcoming the so-called ``fermion sign problem'',  which here manifests in fluctuations of the sign of each of the coefficients $C_i$ and which cannot be pre-determined, depends on the importance of annihilation events among the walkers in establishing the sign structure of the sampled wavefunction. When this effect is strong, the full FCIQMC method requires a number of walkers which scales with the size of the Hilbert space, making it impractical for large spaces. 

To counter this, we use a range of modifications to FCIQMC, which facilitates calculations when the sign problem is strong. We make use of a similarity-transformed Hamiltonian which makes the many-body wave function more compact in Hilbert space \cite{Dobrautz2019}. We also use  the initiator approximation \cite{Cleland2010}, which can introduce a bias into the energy that disappears in the limit of large walker number. In order to control this undesirable bias, we first compare QMC results with exact results in the 1D case and then adjust the walker number until the initiator bias is eliminated. For the 2D systems, we successively increase the walker number with increasing $W$. Walker numbers used in this paper range from $N_W=1\times10^6$ for weakly-interacting 1D chains to $N_W=2\times10^8$ for 2D systems at $U/t=-5$. The most demanding computations were run on up to 400 processor cores using up to 2 Gigabyte memory per core.

\section{Bethe ansatz results}\label{sec:bethe}
For a one-dimensional Hubbard chain ($W=1$), the system is integrable and the Hamiltonian (\ref{eq:hubbard}) can be diagonalized using the Bethe ansatz \cite{Lieb2003}. For a balanced Fermi system with $N$ fermions, energy and total momentum are given by
\begin{align}
\label{eq:BetheEP}
    E&=-2t\sum_{j=1}^N \cos(\kappa_j)+U (L-2N) ,\\
    P/P_0&=\sum_{j=1}^N \kappa_j\quad \text{mod}\quad 2\pi,
\end{align}
with $N$ dimensionless quasi-momenta $\kappa_j$ which must be obtained, alongside $N/2$ rapidities $\Lambda_\alpha$, by solving the Lieb-Wu equations
\begin{align}
\label{eq:LiebWu}
    \exp(i \kappa_j L)&=\prod_{\alpha=1}^{N/2}\frac{\sin \kappa_j-\Lambda_\alpha+iU/4t}{\sin \kappa_j-\Lambda_\alpha-iU/4t} ,\\
    \prod_{j=1}^N\frac{\sin \kappa_j-\Lambda_\beta+iU/4t}{\sin \kappa_j-\Lambda_\beta-iU/4t}&=-\prod_{\alpha=1}^{N/2}\frac{\Lambda_\alpha-\Lambda_\beta+iU/2t}{\Lambda_\alpha-\Lambda_\beta-iU/2t} .
\end{align}
Solving the Lieb-Wu equations via root finding can be done with great accuracy and polynomial effort with particle number. The Bethe ansatz thus provides us with an exact reference for the one-dimensional Hubbard chain.

\begin{figure}[t]
    \centering
    \includegraphics[width=0.49\textwidth]{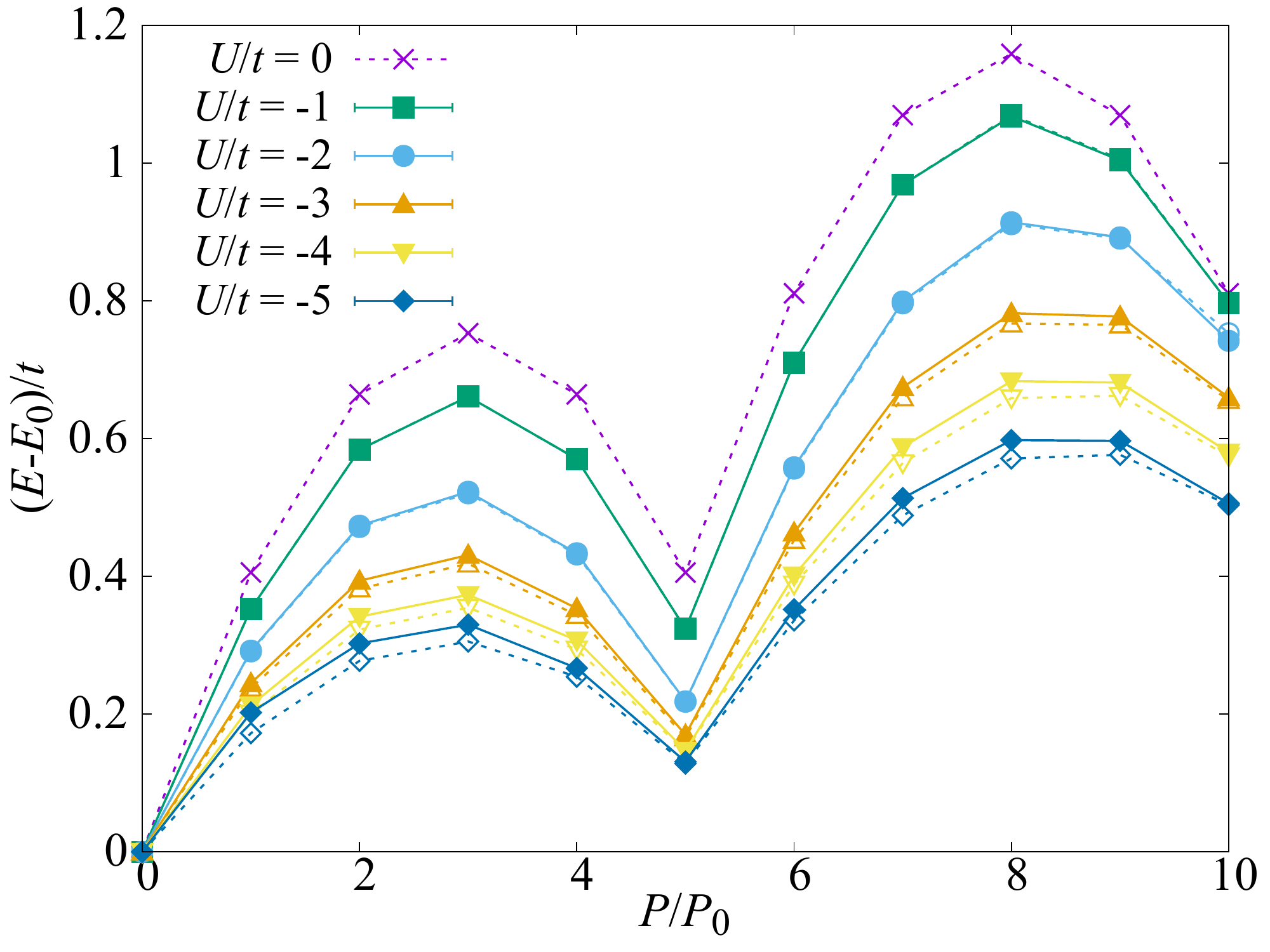}
    \caption{Comparison of yrast state excitation energies for a 1D Hubbard chain of length $L=21$ with $N=10$ fermions, obtained for interaction strengths $U/t=0,-1,-2,-3,-4,-5$ using the Bethe ansatz (empty symbols, dashed lines) and FCIQMC (filled symbols, solid lines). For $|U/t|\le 5$ the agreement is sufficient for reproducing the main features of the yrast dispersion. The lines are merely a guide to the eye.}
    \label{fig:BetheVsFCIQMC}
\end{figure}

This allows us to compare QMC results with exact results to determine the parameter range where we can consider FCIQMC to be reliable in the sense that a possible initiator bias is smaller than the statistical uncertainty inherent in the Monte Carlo approach. In general, larger values of $|U/t|$ lead to stronger correlations in the many-body wave function, which then requires a larger number of Slater determinants to be accurately represented. Also, the fermionic sign problem becomes more severe, which tends to increase the initiator bias in the calculated energy. In FIG.~\ref{fig:BetheVsFCIQMC}, we compare results obtained using the Bethe ansatz and FCIQMC results for a 1D chain ($L=21$, $W=1$). 
We see that for $|U/t|\le 5$ the agreement is very good. We therefore mainly use interaction strengths of $|U/t|\le 5$ in this paper, which covers the entire BEC-BCS crossover and typical values achievable in experiments \cite{Mitra2018}.

\section{Umklapp points} \label{sec:umklapp}
Figure \ref{fig:BetheVsFCIQMC} illustrates the
%The 
characteristic shape of the yrast dispersion, which is 
%depicted in Fig.~\ref{fig:BetheVsFCIQMC}, 
concave downward resembling inverted parabolas in the intervals $0<P<P_0 N/2$ and $P_0 N/2<P<P_0 N$, and with local minima at integer multiples of $P_0 N/2$. It can be understood by looking at the non-interacting case: To increase the total momentum of the system by a single unit of $P_0$, first a fermion at the Fermi surface is excited. The resulting hole can then be filled by another fermion to increase the momentum again, and so forth, with the energy tracing the inverted parabolic part of the Hubbard lattice dispersion relation of Eq.~(\ref{eq:Hubbard_dispersion}). The first local minimum of the yrast spectrum, called the half-umklapp point, in our case ($N=10$) at $P/P_0=5$ is reached when all particles of one spin component have each been boosted by $P_0$. At that point, the Fermi surfaces of both components are shifted with respect to each other, but there are no holes in the Fermi seas of either spin species.

The full umklapp point at $P/P_0=10$ is reached when both spin components, or all particles are boosted. In the continuum limit, where full Galilean invariance is restored, the excitation energy of the umklapp point is determined by the boost only and is independent of interactions, since the state is strictly a boosted ground state. In the lattice system, where Galilean invariance is broken, a weak interaction dependence at the umklapp points is nevertheless observed as seen in Fig.~\ref{fig:BetheVsFCIQMC}.
% Because the entire Fermi sea is boosted, the interaction energy of the full umklapp point is the same as that of the ground state, with the increased .

\begin{figure}
    \centering
    \includegraphics[width=0.49\textwidth]{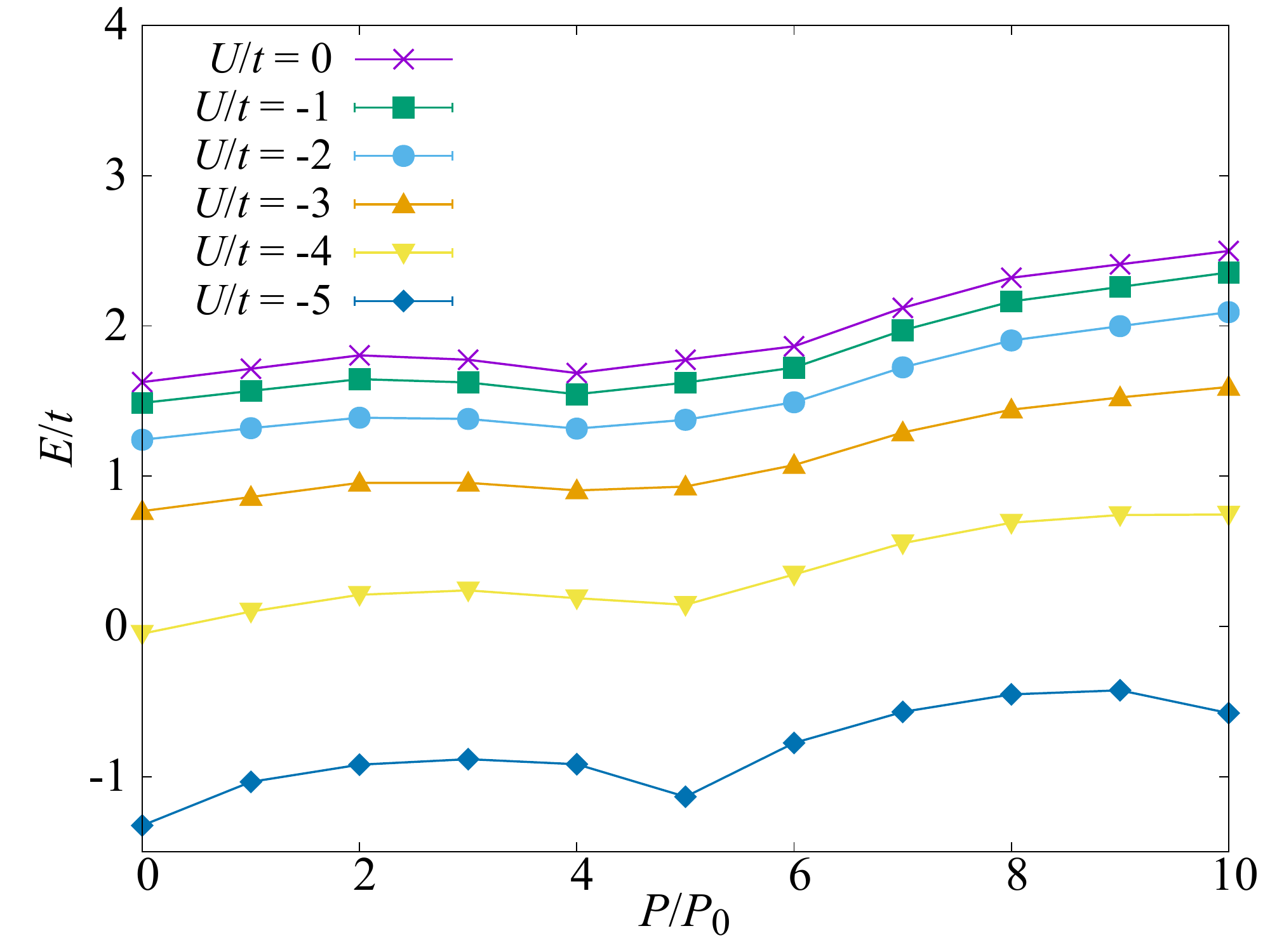}
    \caption{Yrast spectrum for a lattice of $21\times 11$ sites and interaction values $U/t=0,-1,-2,-3,-4,-5$. The spectrum changes from having linear segments to a parabolic shape like in the 1D case depicted in Fig.~\ref{fig:BetheVsFCIQMC}.}
    \label{fig:2DYrastW11}
\end{figure}

For a non-interacting 2D system, constructing the yrast dispersion from hole excitations leads to a different shape, which in the thermodynamic limit in an isotropic 2D system is linear. In our case, as is shown in FIG.~\ref{fig:2DYrastW11} for $W=11$, due to finite-size effects in our mesoscopic system, the spectrum for $U=0$ has linear segments but is not perfectly linear. 
%What is notable here is 
It is remarkable that for increasing interaction strength, the parabolic shape of the 1D spectrum with the umklapp points at $P/P_0=5,10$ is restored.

\begin{figure}
    \centering
    \includegraphics[width=0.49\textwidth]{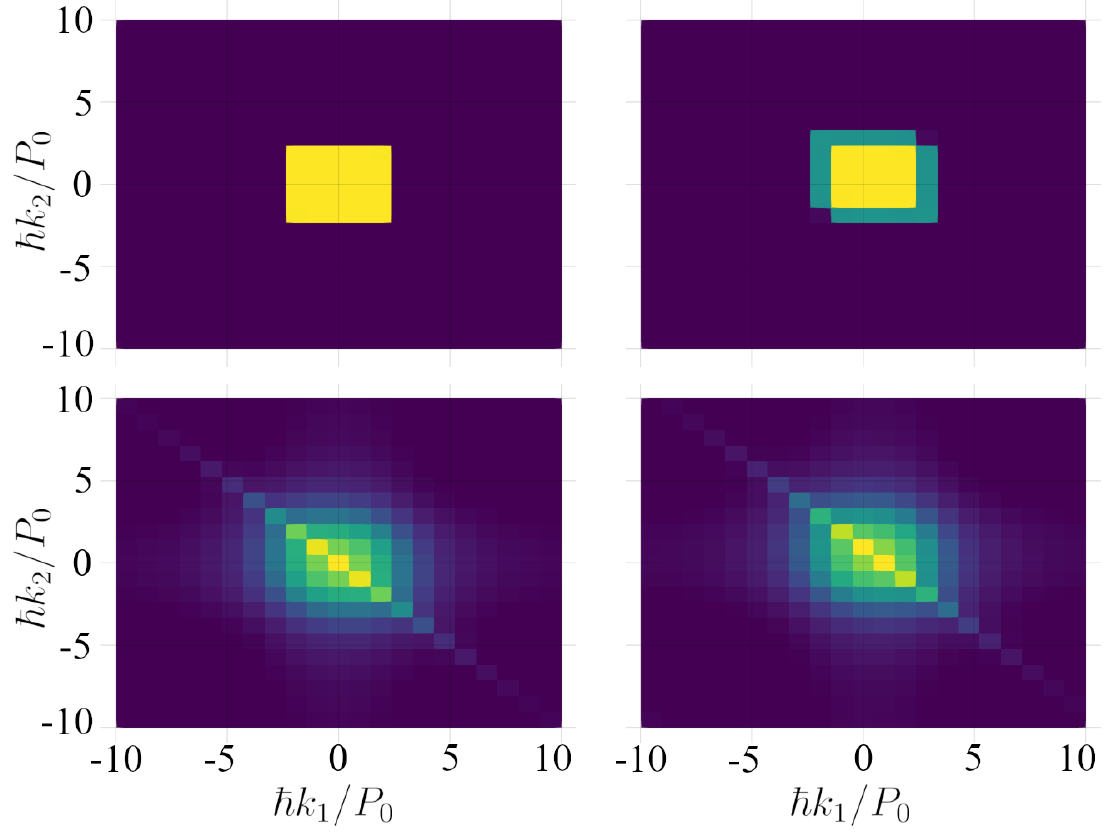}
    \caption{Opposite-spin pair correlation function $g_{\uparrow\downarrow}(k_1,k_2)=\langle c_{k_1,\uparrow}^\dagger c_{k_2,\downarrow}^\dagger c_{k_2,\downarrow} c_{k_1,\uparrow}\rangle$ for the ground state (left) and the half-umklapp point (right) at $U/t=-1$ (top) and $U/t=-5$ (bottom). The 
% BEC-BCS 
crossover from a weakly interacting Fermi gas to a BEC of bound pairs is clearly visible, as the half-umklapp point changes its characteristics from one spin component shifted in momentum space with respect to the other to a translation of the whole system. The half-umklapp point of the Fermi system becomes the first full umklapp point of a fully paired superfluid with 5 bosonic pairs.}
    \label{fig:HalfUmklappCorrelations}
\end{figure}

We can take a look at how the pairing in the system changes the characteristics of the umklapp points by calculating two-body correlation functions. It is possible to obtain the reduced two-body density matrix
\begin{equation}
    \Gamma_{s_1,s_2,s_3,s_4}(\vec k_1,\vec k_2,\vec k_3,\vec k_4)=\langle\Psi_1| c_{\vec k_1,s_1}^\dagger c_{\vec k_2,s_2}^\dagger c_{\vec k_3,s_3} c_{\vec k_4,s_4}|\Psi_2\rangle
\end{equation}
from FCIQMC by simultaneously running two statistically independent QMC simulations with solutions $\Psi_1,\Psi_2$, to avoid biases \cite{Overy2014}. This is valuable even in the 1D case as obtaining the same quantity from the Bethe-ansatz solution is not feasible. To illustrate the BEC-BCS crossover, we show the opposite-spin pair correlation function $g_{\uparrow\downarrow}(\vec k_1,\vec k_2)=\Gamma_{\uparrow\downarrow\downarrow\uparrow}(\vec k_1,\vec k_2,\vec k_2,\vec k_1)$ for the ground state and half-umklapp point in the 1D case in FIG.~\ref{fig:HalfUmklappCorrelations}.

Strong pair correlations with $k_1+k_2=0$ clearly emerge as interaction strength is increased from $U/t=-1$ to $U/t=-5$. 
%Second, the 
The $P/P_0=5$ half-umklapp point at weak interactions exhibits mostly the physics of a non-interacting system, with two Fermi seas displaced with respect to each other. 
%But at
This is in stark contrast to the situation at $U/t=-5$, where the correlation function is the same as for $P=0$ but shifted in momentum space. The mere translation of the pair correlation function is consistent with interpreting the system as superfluid of 5 bosonic pairs, where total momentum $P/P_0=5$ correponds to a full umklapp (i.e.~Galilean boost of the ground state) in contrast to the weakly-interacting Fermi gas, which only reaches a half umklapp point at this momentum.
%, showing 
%that the system now can be considered a gas of 5 bosonic pairs.

\begin{figure}[ht]
    \centering
    \includegraphics[width=0.49\textwidth]{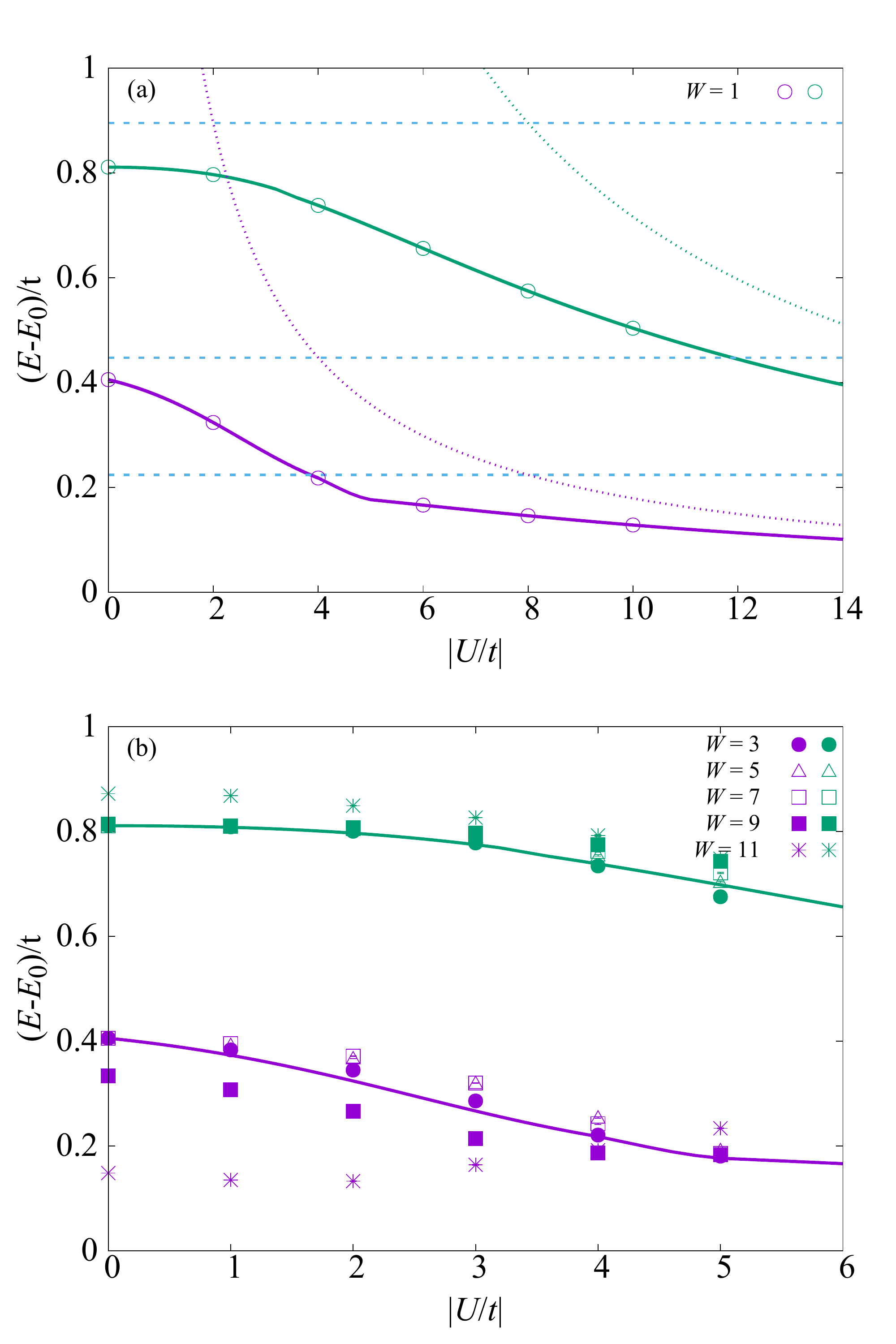}
    \caption{Excitation energy for the half umklapp point $P_h= 5 P_0$  (purple data set at lower energies) and for the full umklapp point $P_u=10 P_0$ (green data set at higher energies) for different values of interaction strength $U$ and width $W$. Solid lines are calculated for $W=1$ using the Bethe ansatz and symbols denote FCIQMC results, for the 1D system in panel (a) and for 2D systems in panel (b). The horizontal dashed lines show the behavior expected in the continuum 1D Fermi gas (YG model) as explained in the main text while the dotted lines show the expected asymptotic behavior for $-U/t \gg 1$. While the excitation energy at the full umklapp point appears to be largely independent of the transverse lattice size indicating a full translation of both Fermi seas in momentum space, the half umklapp point shows additional features of a mesoscopic system for different lattice sizes. Interestingly,  the values for $W=11$ approach the other points from below. For this particular value of $W$, a rearrangement rather than a simple shift of the fermions in momentum space takes place.}
    \label{fig:Umklapp}
\end{figure}

The yrast excitation energy at the half umklapp point $P_h = P_0 N/2$ and at the full umklapp point $P_u = P_0 N$ are shown in Fig.~\ref{fig:Umklapp}. The lattice results can be compared to the expected excitation energies in an equivalent free-space system.
The behavior of the umklapp points of an attractive 1D Fermi gas in free space has been studied using the Yang-Gaudin (YG) model \cite{Shamailov2016}. There, the energy of the full umklapp point is independent of the interaction strength, as it represents simply a translation of the entire Fermi sea in momentum space. In the YG model this energy is given by $E-E_0=P_u^2/2Nm$, where $N$ is the total particle number and $P_u = P_0 N$ the full umklapp momentum. In units of the Hubbard model parameters, this excitation energy is $E-E_0=40\pi^2 t/L^2 $ and is depicted as the upper dashed horizontal line in Fig.~\ref{fig:Umklapp} (a). The half-umklapp point however drops by factor of $2$ from $E-E_0=P_h^2/2mN_\uparrow$ to $E-E_0=P_h^2/2mN$, as it changes from being the half-umklapp point of a system of $N$ fermions to being the full umklapp point of a gas of $N/2$ bosons. This is shown as the two lower dashed horizontal lines in Fig.~\ref{fig:Umklapp} (a), with energies $E-E_0=20\pi^2t/L^2$ and $E-E_0=10\pi^2t/L^2$.

For the non-interacting case we observe expected behavior with energy values slightly lower than the YG model. This is because the YG model uses a quadratic dispersion written in parameters of the Hubbard model as
\begin{equation}
    \epsilon_{\vec k}^\text{YG}=2t\alpha^2{\vec k}^2,
\end{equation}
and $\epsilon_{\vec k} \le \epsilon_{\vec k}^\text{YG} $.

However, for finite interactions the energy values we calculate for the Hubbard model do not behave like for the YG model. In our system, lattice effects dominate once the interaction is strong enough. It is known that for $U/t\rightarrow-\infty$, the asymptotic effective Hamiltonian of the Hubbard model corresponds to a bosonic system with a one-boson-per-site hard-core condition and repulsive next-nearest neighbor interactions \cite{Teubel1990}. The effective Hamiltonian also has a global pre-factor of $U^2/t$, meaning that
% in the asymptotic regime,
the entire spectrum has the same scaling in the asymptotic regime. For the (half-)umklapp points, the asymptotes $80P_0^2/m \times t^2/U$ and $20P_0^2/m\times t^2/U$
%, respectively, 
are presented in Fig.~\ref{fig:Umklapp} (a) as the green and purple dashed lines, respectively. We see that finite-size effects are reduced as the umklapp energies approach these asymptotic lines.

\section{Maxima of the yrast spectrum} \label{sec:maxima}
The point $P = P_h/2$ near the first local maximum of the yrast spectrum (such as at $P/P_0=2,3$ in Fig.~\ref{fig:2DYrastW11}) is of particular interest as a point where in the 1D homogenous case dark solitons appear \cite{Jackson2011,Shamailov2016,Syrwid2018} that are stationary with respect to background and with phase step $\pi$ across the soliton. Dark solitons in a Fermi superfluid are characterized by a localized density depression and a phase jump in the superfluid order parameter around this depression. In a system with periodic boundary conditions, this phase jump must be compensated by a phase gradient along the system, which corresponds to a constant counterflow velocity $v_\text{cf}$. In addition, the soliton can be associated with an inertial mass $m_I$, related to the curvature of the yrast dispersion. We extract these parameters from our calculated dispersions by fitting the quadratic function
\begin{equation}
    E(P)=E(0)+v_\text{cf}(P- P_h/2)+\frac1{2m_I}(P-P_h/2)^2
\end{equation}
around the first local maximum
%into the first inverted parabola 
at momenta $P/P_0 = 1,2,3,4$, where $P_h = NP_0/2 = 5P_0$.

\begin{figure}[ht]
    \centering
    \includegraphics[width=0.49\textwidth]{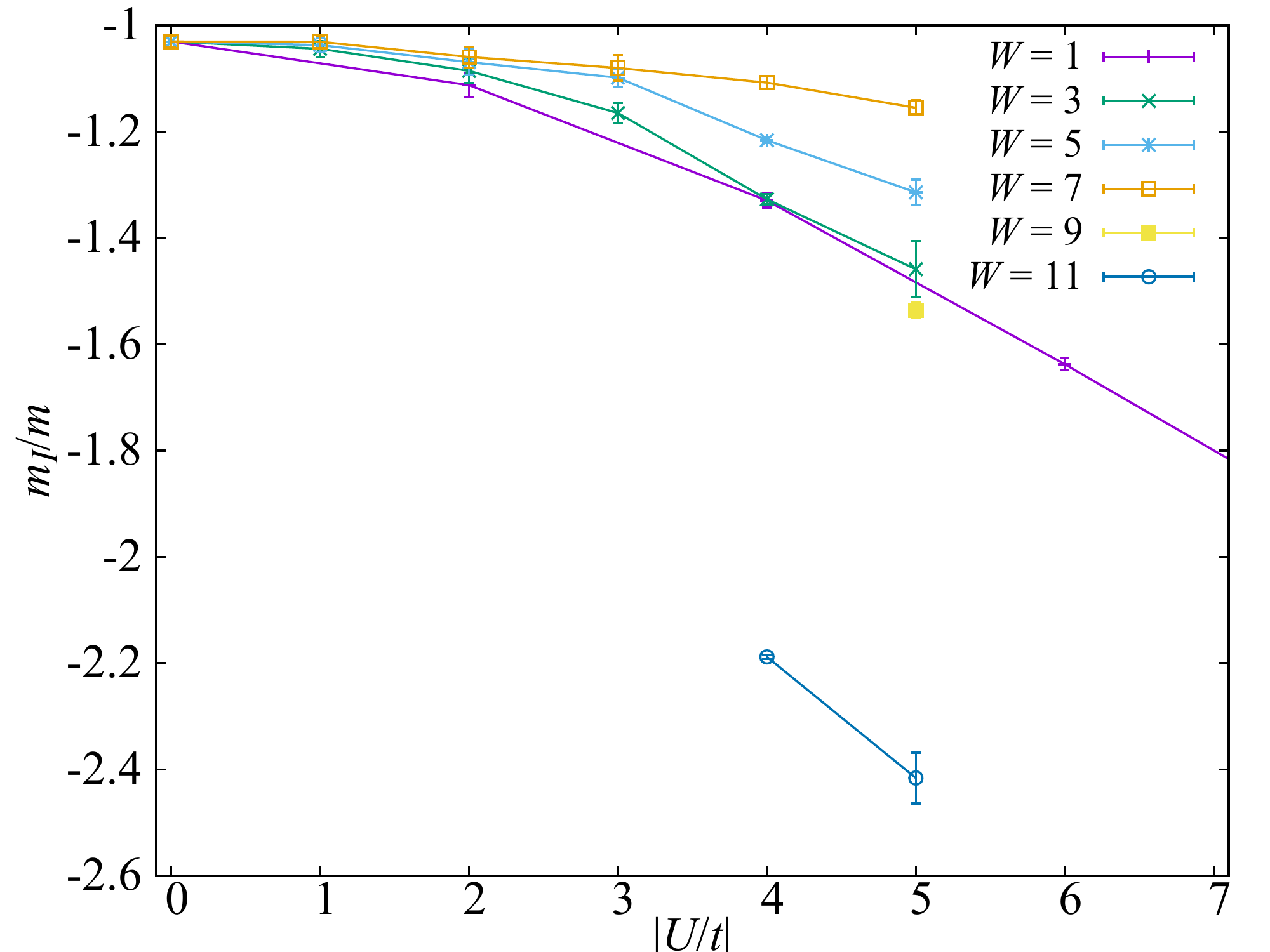}
    \caption{Inertial mass of the first yrast maximum vs.~interaction strength. Unlike in free space, in the lattice, the inertial mass asymptotically approaches $-\infty$ linearly with the interaction strength $U/t$. In the 2D regime, where a parabolic yrast dispersion reappears for strong interactions, the inertial mass is larger by a factor of 2.}
    \label{fig:InertialMass}
\end{figure}

The results for the inertial mass are shown in Fig.~\ref{fig:InertialMass}, where we only show data points for the parameters where the yrast dispersion  closely resembles a parabolic shape. This would correspond to a regime where a superfluid is present and the particles are strongly paired. This is mostly the case for $W\le 7$, where the system is still effectively one-dimensional and transverse momentum states are sparsely populated. For the larger and more 2D systems with $W=7,9$, we find that for $|U/t|=4,5$, a parabolic yrast spectrum reappears (see also Fig.~\ref{fig:2DYrastW11}). For the effective mass, there is an increase in magnitude by a factor of 2 as we increase the system size to $W=11$. This  indicates a further change to the properties of the system, likely a transition from a soliton state to a vortex pair.
This scenario is closely related to the snaking instability of a planar dark soliton in a two-dimensional superfluid, where the soliton decays into pairs
%, which in the time-dependent case would show up as the snaking instability, where a soliton decays into a pair 
of oppositely charged vortices as the system becomes wide enough \cite{Brand2002,Cetoli2013,VanAlphen2019}. However, we cannot directly show a potential density depression caused by the soliton. The reason is that, unlike in mean-field theory, our technique provides the many-body wave function of a (translationally invariant) eigenstate of total momentum and thus a superposition of solitons at all possible positions. The real-space single-particle density we can calculate is flat. To map out the dark soliton as described in \cite{Syrwid2018}, we would need access to higher-order density matrices beyond the two-body density matrix. Therefore, it remains to be seen if the increase in inertial mass really corresponds to a soliton-vortex pair transition.

\begin{figure}[ht]
    \centering
    \includegraphics[width=0.49\textwidth]{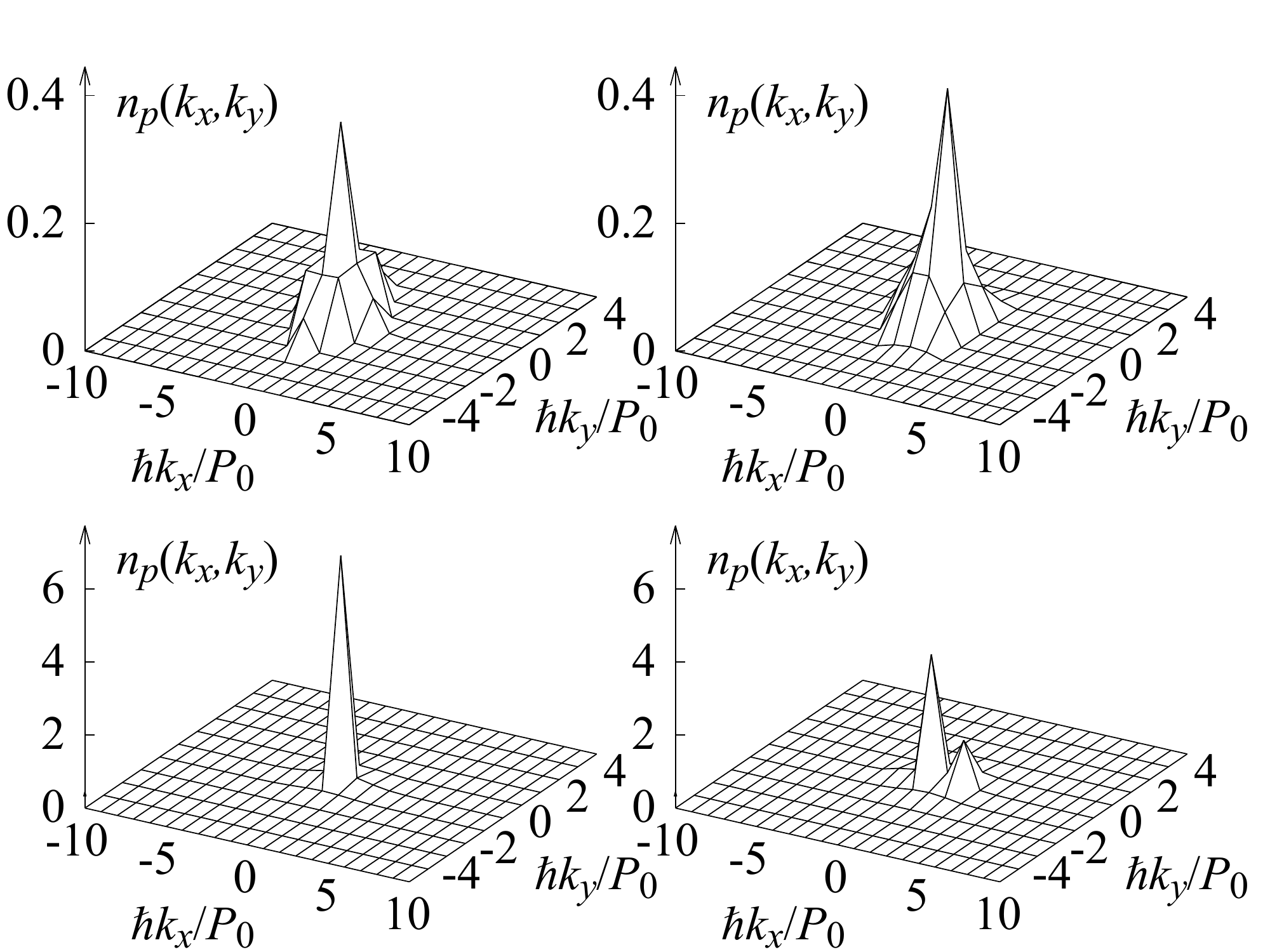}
    \caption{Pair densities for 2D lattices with $21\times11$ sites. The weakly-interacting case ($U/t=-1$, top row) show a small peak in the pair density at zero momentum both for the ground state (top left) and the yrast state at $P/P_0=2$ (top right). For strong interactions ($U/t=-5$, bottom row), the peak at zero momentum dominates at $P=0$ (bottom left), while at $P/P_0=2$ (bottom right), a second peak appears, with the system showing condensation into both zero momentum and momentum $\hbar k_x=2P_0$.}
    \label{fig:2dPairdensity}
\end{figure}

Instead, we investigate more closely the pair condensation,
% in the BEC-BCS transition, 
for which a relevant quantity is the pair Green's function
\begin{equation}
    G_p(l)=\langle \psi_{j+l,\uparrow}^\dagger\psi_{j+l,\downarrow}^\dagger\psi_{j,\downarrow}\psi_{j,\uparrow} \rangle,
\end{equation}
where $\psi$  and $\psi^\dag$ denote creation and annihilation operators, respectively, in position space. The Fourier transform of this Green's function is the momentum-space pair density. For a system with a homogeneous density, it can be directly obtained from the momentum representation of the two-body density matrix
\begin{equation}
    n_p(\vec k)=\sum_{\vec k_1,\vec k_2}\Gamma_{\uparrow\downarrow\downarrow\uparrow}(\vec k_1,\vec k-\vec k_1,\vec k-\vec k_2,\vec k_2).
\end{equation}
This quantity indicates whether Bose-Einstein condensation of pairs occurs. While we find evidence of Bose-Einstein condensation of pairs by a peak in the pair density that 
%a clear peak in the pair density that 
grows with increasing interaction strength for the ground state and umklapp point,
the situation is more complex for general yrast states. 
In Fig.~\ref{fig:2dPairdensity} we show the pair density for a 2D system with $21\times 11$ sites for interaction parameters $U/t=-1$ and $U/t=-5$, for the ground state and the $P/P_0=2$ yrast state. 
For weak interactions, the structure of the pair density is determined mostly by the structure of the Fermi sea, or the noninteracting yrast state.
%In the weakly-interacting case, we see no evidence of strong pairing. 
For stronger interactions, the ground state exhibits one sharp peak at zero momentum indicating strong pairing correlations, as expected for crossover to a BEC of pairs. However, for $P/P_0=2$, there are actually two peaks, for longitudinal momenta $0$ and $2P_0$. Similar features appear in the smaller 2D and 1D systems.

\begin{figure}[ht]
    \centering
    \includegraphics[width=0.49\textwidth]{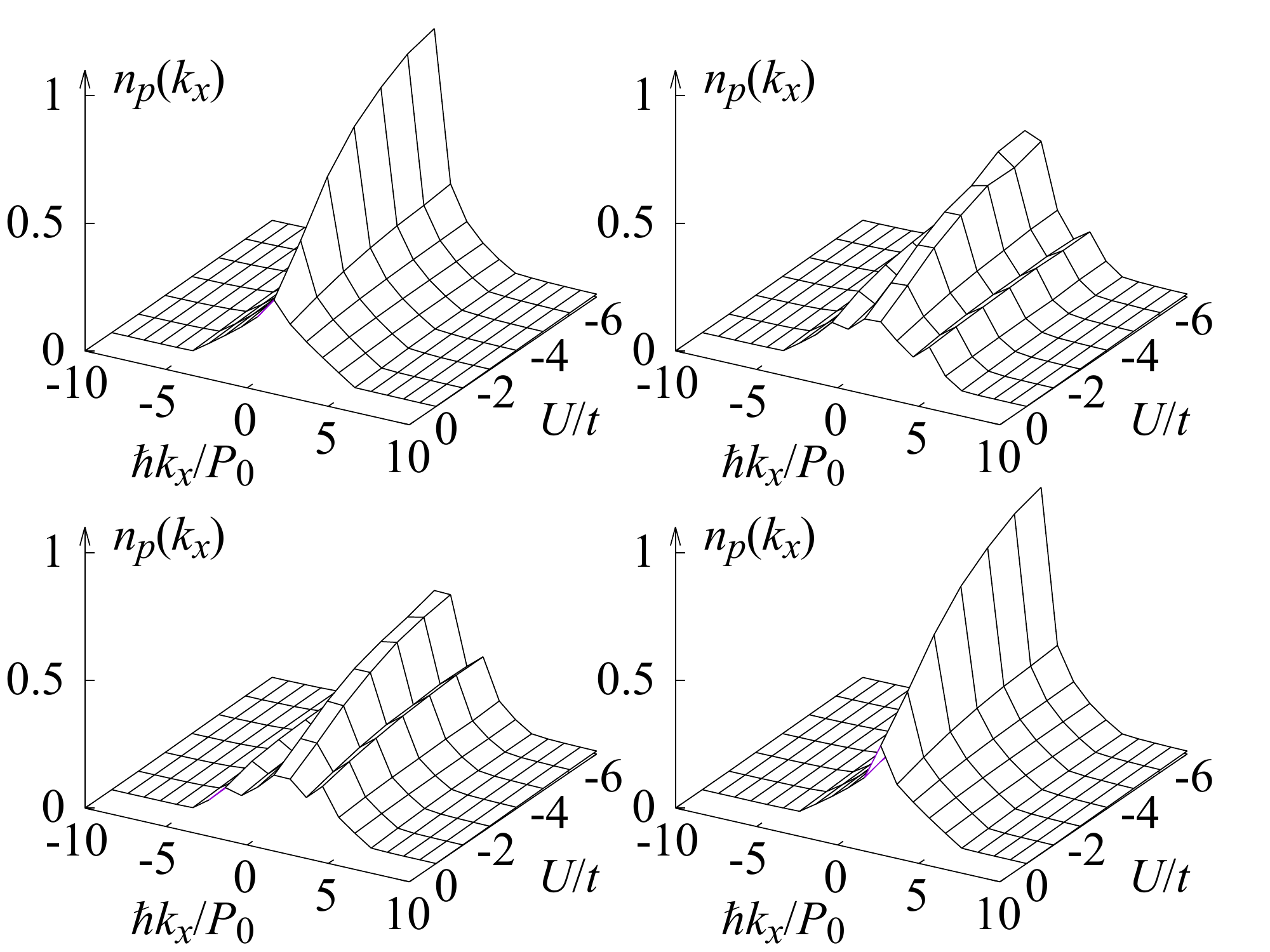}
    \caption{Pair densities for the 1D ground state (top left), the yrast states with $P/P_0=2$ (top right), $P/P_0=3$ (bottom left) and the half-umklapp point (bottom right). The ground state shows a rapid growth of the pair density at zero momentum, as expected for Bose condensation of pairs, while the half-umklapp point is identical to the ground state but shifted by one lattice momentum unit. For the yrast states however, we observe near equal growth of both pair momenta $P=0$ and $P=P_0$, meaning that fragmented condensation into zero and non-zero momentum states is taking place.}
    \label{fig:Fragmentation}
\end{figure}

We now show that this double-peak feature signifies the presence of fragmented condensation. Fragmentation occurs when during the transition to a Bose-Einstein condensate, more than one state becomes macroscopically occupied \cite{Mueller2006}. In Fig.~\ref{fig:2dPairdensity}, the momentum states $(0,0)$ and $(2P_0,0)$ dominate the pair density. For the 1D system, where obtaining the reduced density matrices is easier, we plot the pair densities of several momenta for the yrast state with $P=3P_0$ in Fig.~\ref{fig:Fragmentation} as a function of interaction strength. We see that in addition to $P=0$, the pair density at $P=P_0$ strongly increases as well. Similarly, for other yrast states we see the same phenomenon, a strong signature of fragmented condensation. It is worth noting that the pair density with two peaks obtained here is similar to the case of an imbalanced Fermi gas, where Fermi surfaces of different size lead to FFLO pairing with non-zero total momentum and the signature is a two-peaked pair-density. This has been studied for 1D and 2D Hubbard models  \cite{Wolak2010,Lutchyn2011,Wolak2012,Cheng2018a,Cheng2018b}. Yrast states in our balanced system start with holes in one of the Fermi seas for weak interaction, which also leads to pairing with non-zero total momentum.

\section{Conclusions} \label{sec:conclusions}
In this paper, we have used the FCIQMC method to study the 
%BCS-BEC 
crossover from weakly interacting fermions to a condensate of bosonic pairs
for yrast states in mesoscopic Fermi systems. With this method we can treat larger systems than are accessible to the previously used deterministic CI or exact diagonalization methods
%too large for methods used earlier such as the CI method or exact diagonalization 
\cite{Viefers2004,Manninen2012} and can probe the full transition from a 1D chain not just to quasi-1D rings, but also to full 2D systems. We obtain energy spectra and reduced two-body density matrices for yrast states in the attractive Hubbard model in these geometries.

Comparisons with exact Bethe ansatz results in 1D show very good agreement and demonstrate that FCIQMC can accurately provide yrast states for mesoscopic systems with 10 fermions and $21\times W$ sites where $W$ ranges from 1 to 11. We find that the shape of the yrast spectrum changes from the typical inverted parabolas in 1D to a quasi-linear spectrum in 2D. However, as interaction strength is increased, the parabolic dispersion including the half umklapp point is restored. While the quasi-linear spectrum is a consequence of the 2D geometry and the Pauli exclusion principle, with stronger interactions the exact shape of the non-interacting Fermi sea plays less of a role until we see the expected universal concave downward dispersion of a spinless superfluid.
%behavior of a superfluid of bosonic pairs.
This indicates a transition to a fully paired Fermi superfluid. 

We further find that mesoscopic effects can cause significant deviations for certain geometries from
%while 
the general behavior of the umklapp energies, which otherwise does not differ much between 1D and 2D systems.
%, mesoscopic effects can appear for certain geometries. 
% In our case 
Specifically we find that the half-umklapp excitation energy for a lattice of $21\times 11$ points increases with interaction strength at intermediate values of $U/t$ contrary to the general trend displayed by all other systems under study. 
%has an initially increasing excitation energy. 
This originates in a rearrangement of the fermions in momentum space, where for this particular geometry, the half-umklapp point is a different configuration than the ground state with one spin component shifted in momentum space. This can be of importance for experiments on mesoscopic Fermi systems.

By calculating the pair correlation function, we can follow the pairing process by means of which the fermionic half-umklapp point becomes the first full umklapp point of the Bose condensate of pairs.

In the fully paired regime, we calculated the inertial mass of the dark-soliton-like local maximum of the yrast dispersion. We found a sudden increase by a factor of 2 between the narrow stripe geometry and our largest system in a lattice of $21\times11$ sites. We interpret this as a possible change in the system geometry where a soliton is no longer stable and the yrast state is instead provided by pair of  oppositely charged vortices.

The most striking feature of yrast states in the Hubbard model is revealed to be fragmented condensation, which occurs around the maxima of the yrast dispersion, away from the umklapp points. We find that here, multi-peaked pair densities appear, where in addition to pairing with zero total momentum, the amplitude of other total momentum pairs becomes large for strong interactions. These pair densities of yrast states share some similarities with FFLO states, which are characterized by a double-peaked momentum pair density \cite{Lutchyn2011}. In the FFLO case the origin of this is a mismatch of Fermi surfaces, which have different size due to the spin imbalance. In our case of the yrast states, the Fermi surfaces are shifted in momentum space. In addition, both yrast and FFLO states are related to solitons in the real space density  \cite{Lutchyn2011}.

\begin{acknowledgments}
We thank M.\ Zwierlein, S.\ Shamailov, P.\ Jeszenszki and W.\ Dobrautz for useful discussions. UE  thanks the Max-Planck-Institute for Solid State Research for hospitality.
This work
was supported by the Marsden Fund of New Zealand (Contract No.\ MAU1604), from government funding managed by
the Royal Society Te Ap\=arangi. We also acknowledge support by the NeSI high-performance computing facilities through a Merit allocation and a consultancy project.
\end{acknowledgments}

%\appendix

%\section{Appendixes}
%Appendix...

% The \nocite command causes all entries in a bibliography to be printed out
% whether or not they are actually referenced in the text. This is appropriate
% for the sample file to show the different styles of references, but authors
% most likely will not want to use it.
%\nocite{*}

\bibliography{yrastpaper}

\end{document}